\documentstyle[psfig]{mn}

\newif\ifAMStwofonts
\AMStwofontstrue

\ifoldfss
  \ifCUPmtlplainloaded \else
    \NewTextAlphabet{textbfit} {cmbxti10} {}
    \NewTextAlphabet{textbfss} {cmssbx10} {}
    \NewMathAlphabet{mathbfit} {cmbxti10} {} 
    \NewMathAlphabet{mathbfss} {cmssbx10} {} 
  \fi
  \ifAMStwofonts
    \ifCUPmtlplainloaded \else
      \NewSymbolFont{upmath} {eurm10}
      \NewSymbolFont{AMSa} {msam10}
      \NewMathSymbol{\upi}     {0}{upmath}{19}
      \NewMathSymbol{\umu}     {0}{upmath}{16}
      \NewMathSymbol{\upartial}{0}{upmath}{40}
      \NewMathSymbol{\leqslant}{3}{AMSa}{36}
      \NewMathSymbol{\geqslant}{3}{AMSa}{3E}

    \fi
  \fi
\fi 

\ifnfssone
  \newmathalphabet{\mathit}
  \addtoversion{normal}{\mathit}{cmr}{m}{it}
  \addtoversion{bold}{\mathit}{cmr}{bx}{it}
  \newmathalphabet{\mathbfit} 
  \addtoversion{normal}{\mathbfit}{cmr}{bx}{it}
  \addtoversion{bold}{\mathbfit}{cmr}{bx}{it}
  \newmathalphabet{\mathbfss} 
  \addtoversion{normal}{\mathbfss}{cmss}{bx}{n}
  \addtoversion{bold}{\mathbfss}{cmss}{bx}{n}
  \ifAMStwofonts
    \ifCUPmtlplainloaded \else
      \UseAMStwoboldmath
      \makeatletter
      \new@mathgroup\upmath@group
      \define@mathgroup\mv@normal\upmath@group{eur}{m}{n}
      \define@mathgroup\mv@bold\upmath@group{eur}{b}{n}
      \edef\UPM{\hexnumber\upmath@group}
      \new@mathgroup\amsa@group
      \define@mathgroup\mv@normal\amsa@group{msa}{m}{n}
      \define@mathgroup\mv@bold\amsa@group{msa}{m}{n}
      \edef\AMSa{\hexnumber\amsa@group}
      \makeatother
      \mathchardef\upi="0\UPM19
      \mathchardef\umu="0\UPM16
      \mathchardef\upartial="0\UPM40
      \mathchardef\leqslant="3\AMSa36
      \mathchardef\geqslant="3\AMSa3E
    \fi
  \fi
\fi 

\ifnfsstwo
  \DeclareMathAlphabet{\mathbfit}{OT1}{cmr}{bx}{it}
  \SetMathAlphabet\mathbfit{bold}{OT1}{cmr}{bx}{it}
  \DeclareMathAlphabet{\mathbfss}{OT1}{cmss}{bx}{n}
  \SetMathAlphabet\mathbfss{bold}{OT1}{cmss}{bx}{n}
  \ifAMStwofonts
    \ifCUPmtlplainloaded \else
      \DeclareSymbolFont{UPM}{U}{eur}{m}{n}
      \SetSymbolFont{UPM}{bold}{U}{eur}{b}{n}
      \DeclareSymbolFont{AMSa}{U}{msa}{m}{n}
      \DeclareMathSymbol{\upi}{0}{UPM}{"19}
      \DeclareMathSymbol{\umu}{0}{UPM}{"16}
      \DeclareMathSymbol{\upartial}{0}{UPM}{"40}
      \DeclareMathSymbol{\leqslant}{3}{AMSa}{"36}
      \DeclareMathSymbol{\geqslant}{3}{AMSa}{"3E}
    \fi
  \fi
\fi 

\ifCUPmtlplainloaded \else
  \ifAMStwofonts \else 
    \def\upi{\pi}
    \def\umu{\mu}
    \def\upartial{\partial}
  \fi
\fi

\makeatletter
\newcounter{cureqno}
    {\let\theequation\@curtheeqn%
    \setcounter{equation}{\value{cureqno}}}
\makeatother
\let\gtrsim=\ga
\let\lesssim=\la

\def\Ref    {\item}
\def\eq#1{\begin{equation} #1 \end{equation}}

\def\about  {\hbox{$\sim$}}
\def\ga     {\hbox{$\gtrsim$}}
\def\la     {\hbox{$\lesssim$}}

\def\Lo     {\hbox{$L_\odot$}}
\def\Mo     {\hbox{$M_\odot$}}

\def\tauV   {\hbox{$\tau_V$}}
\def\Mdot   {\hbox{$\dot{\rm M}$}}
\def\days   {\hbox{$^{\rm d}$}}
\def\tyr    {\hbox{${\rm t_{yr}}$}}

\def\mic    {\hbox{$\umu$m}}
\def\micron {\hbox{$\umu$m}}

\def\apss    {\hbox{A\&ASS}}
\def\apj    {\hbox{ApJ}}
\def\apjs   {\hbox{ApJS}}
\def\aj   {\hbox{AJ}}
\def\mnras   {\hbox{MNRAS}}

\def\comm#1 {{\tt (COMMENT: #1)}}
\def\citealt#1 {{\tt (missing ref: #1)}}
\def\citet#1   {{\tt (missing ref: #1)}}
\def\citep#1   {{\tt (missing ref: #1)}}

\pagerange{\pageref{firstpage}--\pageref{lastpage}} \pubyear{2000}

\def\Draft{Version of Sep 21, 2000; submitted to MNRAS}

\title[Mass loss modulations on AGB]
      {100-year Mass Loss Modulations on the Asymptotic Giant Branch}

\author[M. Marengo, \v{Z}. Ivezi\'{c} \& G. R. Knapp]
   {Massimo Marengo$^1$, \v{Z}eljko Ivezi\'{c}$^2$ and Gillian R. Knapp$^2$\\
    $^1$Harvard-Smithsonian Center for Astrophysics, Cambridge, MA 02138;\\
        International School for Advanced Studies, SISSA/ISAS, Trieste, Italy;
        mmarengo@cfa.harvard.edu\\
    $^2$Princeton University, Department of Astrophysical Sciences,
        Princeton, NJ 08544; ivezic,gk@astro.Princeton.edu\\}

\date{\Draft}

\begin{document}

\maketitle

\label{firstpage}

\begin{abstract}
We analyze the differences in infrared circumstellar dust emission between
oxygen rich Mira and non-Mira stars, and find that they are statistically
significant. In particular, we find that these stars segregate in the K-[12]
vs. [12]-[25] color-color diagram, and have distinct properties of the
IRAS LRS spectra, including the peak position of the silicate emission
feature. We show that the infrared emission from the majority of non-Mira
stars cannot be explained within the context of standard steady-state outflow
models.

The models can be altered to fit the data for non-Mira stars by
postulating non-standard optical properties for silicate grains, or by
assuming that the dust temperature at the inner envelope radius is
significantly lower (300-400 K) than typical silicate grain condensation
temperatures (800-1000 K). We argue that the latter is more
probable and provide detailed model fits to the IRAS LRS spectra for
342 stars. These fits imply that 2/3 of non-Mira stars and 1/3 of Mira stars
do not have hot dust ($>$ 500 K) in their envelopes.

The absence of hot dust can be interpreted as a recent (order of 100 yr)
decrease in the mass-loss rate. The distribution of best-fit model
parameters agrees with this interpretation and strongly suggests that
the mass loss resumes on similar time scales. Such a possibility appears
to be supported by a number of spatially resolved observations (e.g. recent
HST images of the multiple shells in the Egg Nebula) and is consistent with
new dynamical models for mass loss on the Asymptotic Giant Branch.
\end{abstract}


\section{INTRODUCTION}

Asymptotic Giant Branch (AGB) stars are surrounded by dusty shells
which emit copious infrared radiation. Even before the IRAS data
became available, it was shown that IR spectra of AGB stars can be
reasonably well modeled by thermal emission from dust with radial
density distribution $\propto r^{-2}$, and optical properties for either
silicate, or carbonaceous grains (Rowan-Robinson \& Harris, 1982, 1983ab).
The availability of 4 IRAS broad-band fluxes and the IRAS LRS database for
thousands of objects made possible more detailed statistical studies.
Van der Veen and Habing (1988) found that AGB stars occupy a
well-defined region in the IRAS 12-25-60 color-color diagram. They also
showed that the source distribution depends on the grain chemistry,
and is correlated with the LRS spectral classification. Bedijn (1987)
showed that in addition to grain chemistry, the most important quantity
which determines the position of a particular source in the IRAS
12-25-60 color-color diagram is its mass-loss rate: more dust, more IR
emission.

A dust density distribution $\propto r^{-2}$ is expected for a
spherical outflow at constant velocity. Since typical
observed velocities are much larger than the estimated escape
velocities, there must be an acceleration mechanism at work.
Early studies by Gilman (1969) and Salpeter (1974ab), and later by
others (e.g. Netzer \& Elitzur, 1993; Habing, Tignon \& Tielens,
1994), showed that the luminosity-to-mass ratios for AGB stars are
sufficiently large to permit outflows driven by radiation pressure.
In this model, hereafter called ``standard'', the dust acceleration is
significant only close to the
inner envelope edge, and thus the dust density is steeper than
$r^{-2}$ only at radii smaller than about several $r_1$. Here
$r_1$ is the inner envelope radius which corresponds to the dust
condensation point. At larger radii the dust density closely follows
the $r^{-2}$ law, in agreement with the observed mid- and far-IR
emission. In particular, Ivezi\'c \& Elitzur (1995, hereafter IE95)
find that such steady-state radiation pressure driven outflow models can
explain the IRAS colors for at least 95\% of dusty AGB stars. In addition to
explaining the IR emission, these models are in good agreement with
constraints implied by independent outflow velocity and mass-loss rate
measurements (Ivezi\'c, Knapp \& Elitzur 1998).

AGB stars are long-period variables (LPV) which show a variety
of light curves. Based on visual light curves, the General Catalog
of Variable Stars (GCVS, Kholopov {\em et al.}, 1985-88) defines regular
variables, or Miras, semiregular (SR) variables, and irregular
variables (L). The distinctive features are the regularity of the light
curves, their amplitudes, and their periods. Some types are further subdivided
based on similar criteria (e.g. SRa, SRb, SRc,...).

AGB stars of different variability types cannot be distinguished
in IRAS color-color diagrams (Habing 1996, and references therein).
This is easily understood as a consequence of the scaling properties
of dust emission (IE95, Ivezi\'c \& Elitzur 1997, hereafter IE97).
Although steady-state models cannot provide a detailed description of variability,
they can be easily augmented when the variability time scale ($\sim$ 1 year)
is shorter than the dust dynamical crossing time through the acceleration zone
($>$ 10 years). In this case the deviations of the dust density profile are minor, 
and the most important effect of variability is the change of
dust optical depth. Optical depth is anticorrelated with luminosity
because of the movement of the dust condensation point: envelopes
are bluer (in all colors) during maximum than minimum light (Le Bertre 1988ab;
Ivezi\'c \& Elitzur 1996, and references therein). Since the change of dust
density profile is negligible, a source cannot leave the track 
in color-color diagrams that corresponds to that density profile and grain
chemistry: during its variability cycle the source simply moves 
along that track due to the change of optical depth. Consequently,
the overall source distribution is not significantly
affected and a random observation cannot reveal any peculiarities:
even if a star is caught during its maximum/minimum light, the only
change is in its somewhat bluer/redder colors (see e.g. the repeated
observations of Mira by Busso et al. 1996). For this reason,
systematic differences in IRAS color-color diagrams for
AGB stars of different variability types are not expected within
the framework of steady-state models.

Kerschbaum, Hron and collaborators (Kerschbaum \& Hron 1992, 1994, 1996,
Kerschbaum, Olofsson, \& Hron 1996, Hron, Aringer, \& Kerschbaum, 1997,
hereafter we refer to these papers as KHc) studied IR emission of
various types of LPVs by combining the IRAS broad-band fluxes with near-IR
observations and the IRAS LRS database. They find differences in observed
properties between Miras and SRb/Lb variables\footnote{According to KHc,
SRa variables appear to be a mixture of two distinct types: Miras and SRb
variables.}: SRb/Lb variables have somewhat higher stellar temperatures
and smaller optical depths than Miras, and can be further divided into
``blue" and ``red" subtypes, the former showing much less evidence for
dust emission than the latter. KHc also show that ``red" SRb/Lb stars have
very similar Galactic scale heights and space densities to Miras, and show a
similar range of IRAS 25-12 color, possibly implying an evolutionary
connection.

Particularly intriguing  are differences between the LRS
spectra of SRb/Lb variables and Miras with ``9.7" $\mu{\rm m}$ silicate emission
(LRS class 2n). Unlike the featureless spectra of carbonaceous grains,
the silicate dust spectra show rich structure (e.g. Little-Marenin \& Little
1988, 1990). This structure may indicate the existence of different dust
species, although some of the proposed classification appears to reflect
the radiative transfer effects (IE95). The peak position of the
``9.7" $\mu{\rm m}$ feature for SRb/Lb sources is shifted longwards, relative
to the peak position for Miras, by 0.2-0.3 $\mu{\rm m}$. In addition,
the ratio of the streng ths of silicate emission features at
18 $\mu{\rm m}$ and 10 $\mu{\rm m}$, $F_{18}/F_{10}$, is larger for SRb/Lb
variables than for Miras. KHc point out that these differences in LRS features
cannot be due to optical depth effects because the feature shapes should resemble
the grain absorption efficiency for the relevant range of optical depths.
Such a correlation between the variability type of a star and the peak
position of its grain absorption efficiency is not expected within the
context of standard steady-state outflow models.

Ivezi\'c \& Knapp (1998, hereafter IK98) compared the distribution of variable
AGB stars in the K-[12] vs. [12]-[25] color-color diagram and found that stars are not
distributed randomly in this diagram, but occupy well defined regions according
to their chemistry and variability type. While discrimination according to
the chemical composition is not surprising, since the optical properties
of silicate and carbon grains are significantly different,
the separation of Miras from SRb/Lb variables is unexpected.

IK98 also show that, while ``standard'' steady-state
models provide excellent fits to the distributions of Miras of all chemical
types, they are incapable of explaining the dust emission from SRb/Lb stars with
silicate dust. They find that the distribution of these stars in the K-[12] vs.
[12]-[25] color-color diagram can be explained by models with dust temperature
at the inner envelope radius significantly lower (300-400 K) than
typical condensation temperatures (800-1000 K). Such an absence of hot dust
for SRb/Lb stars can be interpreted as a recent (order of 100 yr) decrease in the
mass-loss rate. Furthermore, the distribution of these stars in the K-[12] vs. [12]-[25]
color-color diagram implies that the mass-loss rate probably resumes
again, on similar time scales.

The possibility of mass loss changes with short time scales (on the order of
hundred years) seems to be supported by a number of other
observations. Recent HST images of the Egg Nebula obtained by Sahai
{\em et al.} (1997) show concentric shells whose spacing corresponds to a
dynamical time scale of about 100 years. Although there are various
ways to interpret such shells (e.g. binary companion as
proposed by Harpaz, Rappaport \& Soker 1997; instabilities in the
dust-gas coupling proposed by Deguchi 1997), another possible explanation is
dust density modulations due to the mass loss variations on similar time scales.
Time scales of \about 100 years for mass loss variations have also
been inferred for R Hya by Hashimoto et al. (1998) and for IRC+10216 by
Mauron \& Huggins (1999), and a somewhat longer scale for $\mu$ Cep by
Mauron (1997). The discovery of multiple CO
winds reported by Knapp {\em et al.} (1998) is also in agreement with the
hypothesis of variable mass-loss.

Mass-loss rate modulations with time scales of the order 100 years are
difficult to explain in terms of the periodicities characteristic for AGB stars.
Periodic flashes of the He burning shell in the thermally pulsing AGB phase (known
as ``thermal pulses'', TP hereafter) occur on
timescales of $10^3$--$10^4$ yr (see e.g. Speck, Meixner \& Knapp 2000).
On the other end, the typical pulsational
periods of AGB stars (50 -- 500 days) are too short. Various attempts to
solve this puzzle have been proposed, and are well described by Sahai
{\em et al.} (1998). The basic ideas involve either long period modulations
of the luminosity variations, or temperature fluctuations caused by giant
convection cells in the AGB atmosphere. However, similar mass-loss rate
modulations are also reproduced, without any such additional assumptions, within
the framework of time dependent wind models with driving force varying on time
scales of \about 1 yr (Winters 1998, see also \S 4).

Motivated by these results, we further analyze correlations between
variability type and IR emission for AGB stars, especially the
properties of LRS data, and their implications for the models of steady-state
radiatively driven outflows. In addition to the silicate peak position,
we also use synthetic colors calculated from LRS data to study detailed
properties of mid-infrared emission. Earlier findings that Miras and SRb/Lb
variables exhibit significantly different IR emission properties are confirmed
at statistically significant levels in \S 2. We attempt to explain these
discrepancies by an interrupted mass loss model, and provide detailed model testing
in \S 3. In section \S4 we discuss the implications.

\section{OBSERVED DIFFERENCES IN DUST EMISSION BETWEEN MIRA AND SR STARS}

\subsection{Distribution of AGB stars in K-[12] vs. [12]-[25] Color-color Diagram}

\begin{figure*}
\begin{minipage}{\textwidth}
\label{K1225}
\centering \leavevmode \psfig{file=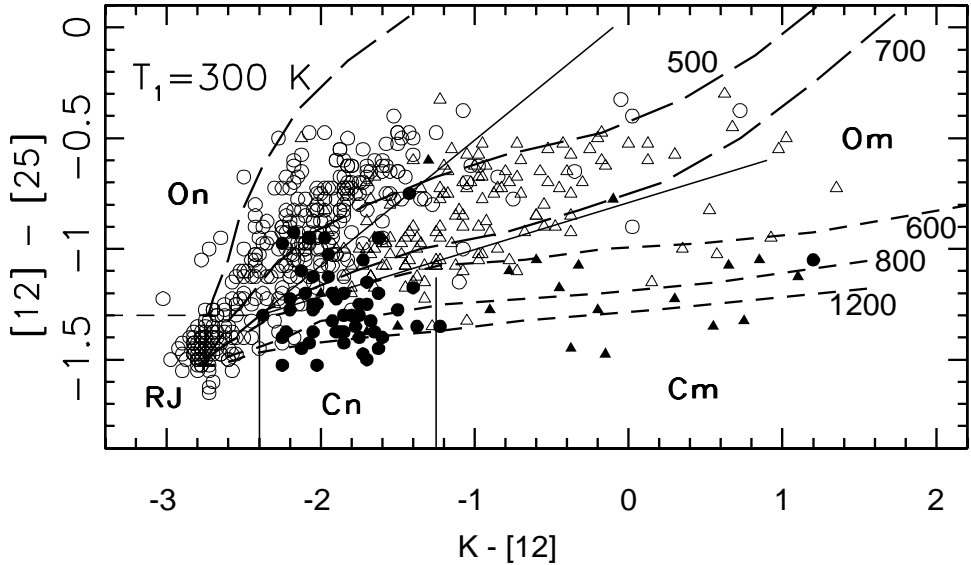,width=\hsize,clip=}
\caption{[12]-[25] vs. K-[12] color-color diagram for a sample of AGB
stars discussed in IK98. Oxygen rich Miras and SRb/Lb stars are marked by
open triangles and circles, respectively. Carbon rich stars are analogously
marked by filled symbols. The thin straight solid lines display a scheme which
classifies stars into 4 groups according to their variability type (Miras, m,
vs. non-Miras, n) and chemistry (O vs. C). The RJ label marks the locus of 
dust-free stars. The thick curved short-dashed and
long-dashed lines represent tracks for interrupted mass-loss models which are
decribed in detail in \S 3. It is assumed that mass loss abruptly stops and
the envelope freely expands thereafter. Note that the models imply a lack of
hot dust for oxygen rich SRb/Lb stars.}
\end{minipage}
\end{figure*}

Figure 1 shows the K-[12] vs. [12]-[25] color-color diagram taken from
IK98\footnote{Note that color definitions are different than in IK98, here we
define all colors as [2]-[1]=-2.5log(F$_2$/F$_1$), where F$_1$ and F$_2$ are
fluxes in Jy.}
(the K data were obtained by Franz Kerschbaum and collaborators). 
The thick curved short-dashed and long-dashed lines represent tracks for
interrupted mass-loss models which are described in detail in \S 3. It is
assumed that mass loss abruptly stops and the envelope freely expands
thereafter. Due to this expansion the dust temperature at the inner envelope radius
decreases. Model tracks for various values ranging from 1200 K to 600 K for carbon
grains (short-dashed lines) and from 700 K to 300 K for silicate grains (long-dashed
lines) are shown in the figure. These tracks demonstrate that the distribution of
stars with silicate dust in K-[12] vs. [12]-[25] color-color diagram can be
reproduced by varying T$_1$ in the range 300--700 K, without any change in the
adopted absorption efficiency for silicate grains. The distribution of Mira
and non-Mira stars with carbon dust is fairly well described by models with T$_1$
= 1200 K, and thus the assumption of significantly lower T$_1$ is not required
by the data. However, note that such models are not ruled out by the data
since all carbon dust models produce similar tracks.

\subsection{IRAS LRS Differences between Oxygen Rich Mira and SR Stars}

The oxygen rich stars usually show infrared spectra indicating silicate dust.
Due to many features in their LRS spectra, such stars are especially well suited
for a comparative study of various subclasses. For this reason, in the rest of this
work we limit our analysis to oxygen rich stars.
The stars without IRAS LRS data are excluded and the final list includes 342 sources,
consisting of 96 Mira stars, 48 SRa, 140 SRb and 58 Lb stars. According to Jura \& Kleinmann
(1992) and Kerschbaum \& Hron (1992), the evolutionary state of the SR variables can
be determined from their pulsation period, $P$: stars with $P < 100^d$
should be in the Early-AGB phase, while stars with $P > 100^d$ are in the
TP-AGB phase. It has been suggested that only stars in the TP-AGB phase suffer
significant mass loss (e.g. Vassiliadis \& Wood 1993). 
We have cross-correlated this list with the General Catalogue of
Variable Stars (GCVS, Kholopov {\em et al.} 1988) and extracted the pulsational period
for SR stars, when available. We find 57 stars with $P < 100^d$ and 130 with
$P \ga 100^d$. 

We divide the stars from the list into several subsamples and analyze the
properties of each group separately:

\begin{itemize}
\item
Mira stars
\item
Non-Mira stars, including SRa, SRb and Lb types, hereafter NM
\item
SRa stars, in order to test the possibility that SRa are a spurious class
containing a mixture of Mira and SRb variables.
\item
Short-period ($P < 100^d$) SR stars, hereafter spSR
\item
Long-period ($P > 100^d$) SR stars, hereafter lpSR
\end{itemize}

Due to the heterogeneous nature of the samples from which the KHc list
is derived, and to the limitations of the IRAS catalog (e.g. source confusion
in the galactic plane), our sample is not statistically complete and may
have hidden biases. However, it is sufficiently large to allow a critical
study of the correlation between variability type and properties of infrared
emission for oxygen rich galactic AGB stars.

\subsubsection{               Mid-IR colors           }

The LRS spectra for a large number of sources can be efficiently compared
by using synthetic colors based on ``fluxes" obtained by convolving LRS spectra
with suitably defined narrow filters. We utilize the photometric system
developed by Marengo {\em et al.} (1999, hereafter M99) as a diagnostic tool
for mid-IR imaging of AGB stars. The system is designed to be sensitive to
the strength of the 9.7 \mic\ silicate feature and to the slope of the dust
continuum emission in the 7--23 \mic\ spectral range. We calculate synthetic
fluxes from LRS data by using 10\% passband filters with Gaussian profiles
centered at 8.5 \mic, 12.5 \mic, and 18.0 \mic. The resulting [8.5]-[12.5] vs.
[12.5]-[18.0] color-color diagram for all stars in the sample is shown
in Figure 2. Mira stars are shown as solid circles, NM stars as open circles,
and SRa stars as gray dots. The line shows the locus of black-body colors
parameterized by the temperature, as marked in the figure.

The differences between Mira and NM sources are evident. Mira stars are
grouped around the black body track, with $-0.1 \lesssim$ [8.5]-[12.5]
$\lesssim 0.4$. This region, as shown by M99, is well described by envelopes
with hot dust and intermediate optical depth $1 \lesssim \tau_V \lesssim
30$. All NM stars have instead a much larger spread in the [8.5]-[12.5]
color, and a redder [12.5]-[18.0] color. The SRa stars have a distribution
similar to other NM stars, except that their [8.5]-[12.5] color
is on average redder than for the whole sample, and in the range observed
for Mira stars.

We provide a quantitative test of the Mira/NM separation in the [12.5]-[18.0]
vs. [8.5]-[12.5] color-color diagram in Figure 3, and summarize it in
Table 1. The histogram in Figure 3 shows the distributions of the [12.5]-[18.0]
color excess, defined as the difference between [12.5]-[18.0] color and the color
of a black body with the same [8.5]-[12.5] color temperature\footnote{Note that
the black body assumption is not crucial since it simply provides a reference
point.}, for Mira, SRa and NM stars. The mean value of this excess is $\sim 0.11$
for Miras, $\sim 0.32$ for NM and $\sim 0.33$ for SRa. The 0.2 magnitude
difference between the mean values of Mira and NM distributions
is slightly larger than the dispersion of the two samples, as measured by
the sample variance ($\sigma \sim 0.18$ for both Miras and NM). We tested this
result with a Student's t-test for the mean values, finding that the difference
between the two populations is statistically significant (see Table 1 for
details). We also performed the same analysis for the spSR and lpSR subsamples
and found that they have a similar mean color excess (0.36 and 0.30 mag);
the difference of 0.06 mag is not significant.

\begin{figure}
\label{MMcc}
\centering \leavevmode \psfig{file=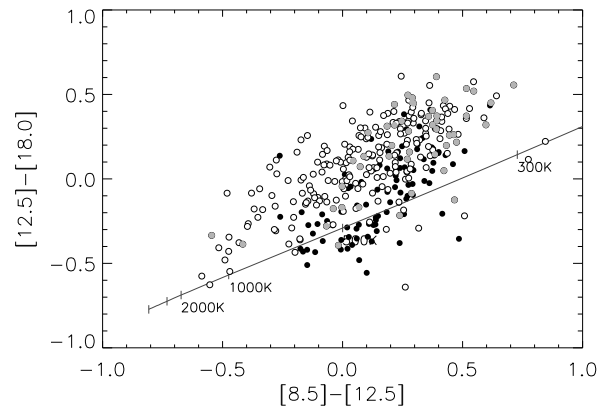,width=\hsize,clip=}
\caption{The [12.5]-[18.0] vs. [8.5]-[12.5] color-color diagram based
on synthetic fluxes obtained from LRS data (see \S2.2.1). Mira stars are
shown as solid circles, NM (non-Mira) stars as open circles, and SRa stars as
gray dots. The line is the locus of black-body colors parametrized by the
temperature, as marked. Note the different distribution of Mira and NM stars.}
\end{figure}

\begin{figure}
\label{colhist}
\centering \leavevmode \psfig{file=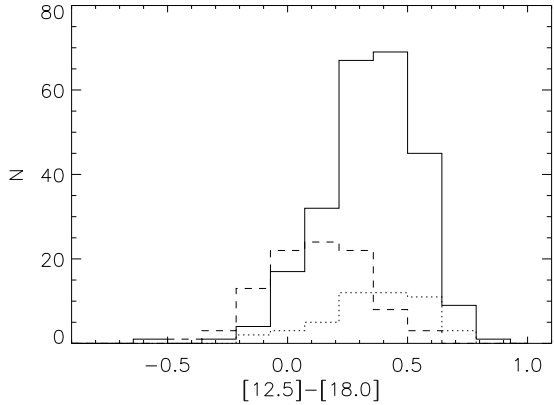,width=\hsize,clip=}
\caption{Distribution of the [12.5]-[18.0] color excess, relative to a black
body with the same [8.5]-[12.5] color temperature, for Mira
(dashed line), NM (solid line) and SRa (dotted line) stars. Non-Mira stars
have \about 0.2 mag larger color excess than Mira stars (see Table 1).}
\end{figure}

\begin{table}
\begin{center}

\begin{tabular}{lcccr}
\hline
 Sample & Median & Mean & $\sigma$ & N \\
\hline \hline
Mira                      & 0.10 & 0.11 & 0.18 &  96 \\
NM                        & 0.34 & 0.32 & 0.18 & 246 \\
SRa                       & 0.37 & 0.33 & 0.19 &  48 \\
spSR ($P < 100$\days)       & 0.37 & 0.36 & 0.17 &  57 \\
lpSR ($P \gtrsim 100$\days) & 0.32 & 0.30 & 0.20 & 130 \\
   \hline
\end{tabular}

\begin{tabular}{lcccr}
\hline
 Sample & F-test & t-test & result \\
\hline \hline
Mira vs. NM    & 0.90   & $6 \cdot 10^{-20}$
                        & Same $\sigma$, diff. mean   \\
 SRa vs. Mira  & 0.70   & $2 \cdot 10^{-10}$
                        & Same $\sigma$, diff. mean   \\
 SRa vs. NM    & 0.60   & 0.71  & Same $\sigma$ and mean     \\
spSR vs. lpSR  & 0.16   & 0.05  & Same $\sigma$ and mean     \\
\hline
\end{tabular}

\caption{Significance of statistical tests for the [12.5]-[18.0] color excess.
Samples have different $\sigma$ at the 99\% confidence level if the variance
F-test returns a significance of 0.01 or smaller, and different mean values if
Student's t-test returns a significance of 0.01 or smaller}
\smallskip

\end{center}
\end{table}

These two statistical tests suggest that, regarding their mid-IR [8.5]-[12.5]
and [12.5]-[18.0] colors, Mira stars and NM stars are different.
That is, given the envelope optical depth, as measured by the [8.5]-[12.5]
color, NM stars show more cold emission measured by the [12.5]-[18.0] color.
Among the SR sources, the mid-IR colors do not correlate with the pulsational
period. Assuming the validity of the Kerschbaum \& Hron (1992) correlation between
the pulsational period and AGB evolution, this suggests that SR stars form
similar circumstellar envelopes in the early and thermally pulsing AGB phases.
The similarity of SRa with NM in general indicates that the SRa sample
is not significantly contaminated by Mira stars.

\subsubsection{The peak wavelength of the ``9.7 \mic" silicate feature}

Following KHc, we also determine the peak wavelength of the ``9.7 \mic"silicate
feature for all sources in the sample. Such studies were done first by
Little-Marenin \& Little (1990) who tried to classify LRS spectra for a large sample
of AGB stars. They found that their subsample of SR and L variables
showed a narrower silicate feature, and shifted to the red compared to
Mira stars. A similar analysis was performed by Hron, Aringer \& Kerschbaum (1997)
for a larger number of sources, accounting for the dust continuum emission by
fitting it with a separate black body. They confirmed the differences between
the two classes, finding a shift of about 0.3 \mic\ in the silicate feature
peak position.

\begin{figure}
\label{peakpos}
\centering \leavevmode \psfig{file=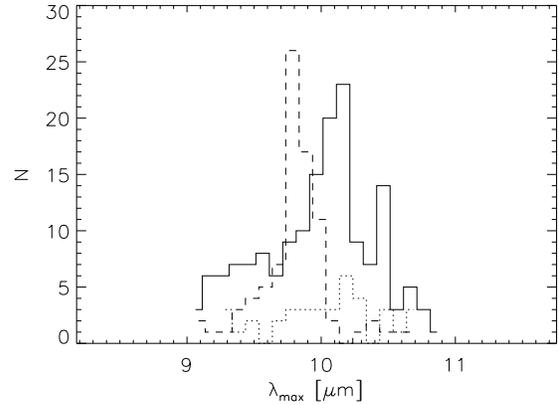,width=\hsize,clip=}
\caption{Distribution of the peak position of the silicate ``9.7" \mic\
feature for Mira (dashed line), NM (solid line) and SRa (dotted line) stars.
The NM stars have the feature shifted for \about 0.2 \mic\ with respect
to the Miras.}
\end{figure}

We determine the silicate feature peak position by fitting a fifth degree
polynomial to each LRS spectrum in the 9--11 \mic\ wavelength range. This
fitting procedure removes the effect of noise and eventual secondary
features, producing ``smooth'' spectra where the position of the maximum
can be easily recognized. The wavelength of the maximum of the fitting
polynomial in the given interval is then assumed to be the position of
the ``true'' silicate peak. With this technique we measured the position of
the silicate peak for 85 Miras and 162 NM; for the remaining sources
(11 Miras and 84 NM) the fitting procedure failed either because of the
excessive noise in the spectra, or due to the absence of a significant feature.
Many of the sources for which the silicate feature is too weak to measure
the position of its peak are Lb stars of LRS class 1n.

The results for the position of the silicate feature peak are listed in
Table 2 and shown as histograms in Figure 4. The mean values for Miras and
NM are $\langle \lambda_{max} \rangle_{Mira} \simeq 9.86$ \micron~ and
$\langle \lambda_{max} \rangle_{NM} \simeq 10.01$ \micron~ respectively;
the medians are 9.87 and 10.08 \micron. Note the different standard deviation
$\sigma_\lambda$ of the two samples, almost twice as large for NM
($\sim 0.42$ \micron) than for Miras ($\sim 0.28$ \micron); this
difference is 99.9\% statistically significant according to the
F-test. The Student's t-test for the mean values,
performed between the two samples (with unequal variances), confirms the
separation of the Mira and NM populations with respect to their
$\lambda_{max}$ at a high (99.95\%) significance level, in agreement with
the results of Hron, Aringer \& Kerschbaum (1997).

\begin{table}
\begin{center}

\begin{tabular}{lcccr}
\hline
 Sample & Median & Mean & $\sigma$ & N \\
\hline \hline
Mira                        &    9.87  &     9.86 & 0.28 &     85 \\
NM                          &   10.07  &    10.01 & 0.42 &    162 \\
SRa                         &   10.10  &    10.08 & 0.41 &     37 \\
spSR ($P < 100$\days)       &   10.09  &    10.04 & 0.43 &     31 \\
lpSR ($P \gtrsim 100$\days) &   10.07  &    10.02 & 0.38 &     95 \\
   \hline
\end{tabular}

\begin{tabular}{lcccr}
\hline
 Sample & F-test & t-test & result \\
\hline \hline
Mira vs. NM   & $5 \cdot 10^{-5}$ & $5 \cdot 10^{-4}$
              & Diff. $\sigma$ and mean$^b$           \\
SRa vs. Mira  & $6 \cdot 10^{-3}$ & $4 \cdot 10^{-3}$
              & Diff. $\sigma$ and mean$^b$           \\
SRa vs. NM    & 0.82   & 0.43  & Same $\sigma$ and mean   \\
spSR vs. lpSR & 0.41   & 0.87  & Same $\sigma$ and mean   \\
\hline
\end{tabular}

\caption{Significance of statistical tests for the peak Position of the
silicate 9.7 \mic\ feature.
Samples have different $\sigma$ at the 99\% confidence level if the variance
F-test returns a significance of 0.01 or smaller, and different mean values if
Student's t-test returns a significance of 0.01 or smaller}
\smallskip

\end{center}
\end{table}

We performed the same test on the SRa class alone, finding $\langle
\lambda_{max} \rangle_{SRa} \simeq 10.08$ \micron, and $\sigma_\lambda
\simeq 0.41$, consistent with the results for the whole NM class. This
further indicates that the SRa subsample does not contain
a significant number of misclassified Miras. The $\langle \lambda_{max} \rangle$
values for the NM subsets with $P \gtrsim 100$ \days~ and $P < 100$ \days~
are 10.07 and 10.09 \micron~ respectively, and the $\sigma_\lambda$ are 0.43
and 0.39 \micron. The difference in the $\sigma_\lambda$ of the two
subsample is not significant, confirming that the two subsets are
part of the same population, in agreement with the analysis of
synthetic mid-IR colors.

\subsection  {   Interpretation of the Observed Differences   }

We have shown that oxygen rich Mira and non-Mira stars have statistically
different properties of their infrared emission. In particular, we find

\begin{enumerate}
\item
Different distribution in the K-[12] vs. [12]-[25] color-color diagram.
\item
Different [12.5]-[18.0] excess emission
\item
Different peak position of the silicate feature
\end{enumerate}

As shown by IK98 (c.f. Figure 1), these differences are not
in agreement with the ``standard" steady-state radiatively driven
wind models: while they can reproduce the properties of the Mira
sample, they cannot explain the infrared emission from non-Mira stars.
There are three ways to augment the ``standard" models without
abandoning the hypothesis of a radiatively driven outflow:
changing the input (stellar) spectrum, altering the spectral
shape of dust opacity, and employing a different dust density
distribution.

The star in the ``standard" models is assumed to radiate as a 2500 K
black-body. The observed differences in the K-[12] vs. [12]-[25] color-color
diagram could imply that typical stellar temperature is significantly
different for Mira vs. non-Mira stars, or that black-body spectral shape
is not an adequate description of the true stellar spectrum (e.g.
due to an absorption feature in K band like H$_2$O feature
seen in IRTS data by Matsuura et al. 1998). A different input spectrum
would affect the model K flux, and to some extent the 12 $\mu{\rm m}$ flux
in optically thin envelopes. However, such a change cannot account
for the observed differences in the mid-infrared region (items 2 and 3
above) where the flux is dominated by dust emission.

It is possible to produce model spectra in agreement with data
for non-Miras by altering the adopted spectral shape of absorption
efficiency for silicate grains. There are two required changes.
First, the ratio of silicate feature strengths at 18 $\mu{\rm m}$ and 10
$\mu{\rm m}$ has to be increased for a factor of about 2-3 \footnote{This
strength ratio is not a very constrained dust property, but the uncertainty
seems to be smaller than a factor of 2 (Draine \& Lee, 1984)}. Such a change
results in a model track in the K-[12] vs. [12]-[25] diagram which leaves the
Rayleigh-Jeans
point at a larger angle (with respect to K-[12] axis) than before, and passes
through the observed source distribution. These models also produce larger
[12.5]-[18.0] excess emission in agreement with item 2 above. In order to
account for item 3 above, a second change is required: the peak position of
the ``9.7" micron feature has to be shifted longwards for about 0.2-0.3 $\mu{\rm m}$.
For example, the inclusion of Al oxides can shift the peak position to longer 
wavelengths (Speck {\em et al.} 2000). We conclude that, although the postulated 
changes might seem ad hoc and have little support from model dust optical 
properties, the possibility of non-Miras having somewhat different grains 
cannot be ruled out by using near-IR and IRAS data alone\footnote{Indeed, 
Sloan {\em et al.} (1996) found that the ``13 micron'' feature occurs somewhat 
more frequently in SRb stars (75\%-90\%) than in all AGB stars with silicate 
dust (40\%-50\%).}.

The third way to alter the model spectra predicted by ``standard"
models is to decrease T$_1$, the dust temperature at the inner
envelope edge, which is usually assumed to correspond to the dust condensation
temperature (\about 1000 K). The removal of hot dust reduces the flux at
12 $\mu{\rm m}$ significantly more than the fluxes in K band and at
25 $\mu{\rm m}$, and produces model tracks in agreement with the
source distribution in the K-[12] vs. [12]-[25] color-color diagrams
(c.f. Figure 1), and larger [12.5]-[18.0] excess emission. It is remarkable
that the models with a low $T_1$ are also capable of
explaining the differences between the peak position of the silicate feature.
The top panel in Figure 5 displays two model spectra obtained with visual optical
depth of 0.6, and T$_1$ = 300 K (dashed line) and 700 K (solid line). The dotted
line shows the stellar spectrum. Because the hot dust has been removed
in the model with
T$_1$ = 300 K, both the blue and red edges of the 10 $\mu{\rm m}$
silicate emission
feature are shifted longwards for about 0.2-0.3 $\mu{\rm m}$. While the peak position
in the model spectra is exactly the same in both models, the addition of noise
to the models would make the whole feature for T$_1$ = 300 K model appear shifted
longwards. Thus, assuming a lower T$_1$ in models for NM stars can
account for all 3 items listed above. Such models are further described in
the next section.

\begin{figure}
\label{Teffect}
\centering \leavevmode \psfig{file=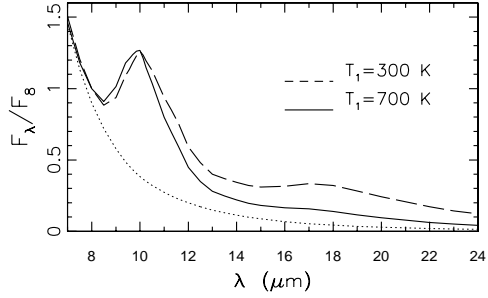,width=\hsize,clip=}
\caption{Model spectra in the LRS region obtained for silicate grains
with the visual optical depth of 0.6 and 2 values of the dust temperatures at
the inner envelope radius, T$_1$ (solid and dashed lines). Dotted line shows
the stellar contribution to the spectra. Note that the silicate ``9.7" \mic\
feature appears weaker and is shifted longwards for T$_1$ = 300 K, relative
to the position obtained for T$_1$ = 700 K, although both are
calculated with the same grain optical properties.}
\end{figure}

We conclude that the two most probable explanations for the observed
differences in infrared emission between Mira and non-Mira stars
are either different grain optical properties, or the lack of hot
dust in non-Mira stars. The possibility that employing different types
of grains could reproduce the observed differences between Mira and
NM stars requires extensive modeling effort and will be analyzed in a
separate publication (Helva\c ci {\em et al.} 2001). Here we discuss low-T$_1$
models and show that they are consistent with the infrared emission observed
for NM stars. Various observational techniques for distinguishing these two
hypothesis are further discussed in \S4.

\section{                       MODELS                  }

\subsection  {      Model Assumptions and Predictions        }

We calculate model spectra by using the DUSTY code (Ivezi\' c,
Nenkova \& Elitzur, 1997). It is assumed that the star radiates as a
2500 K black-body. Silicate warm dust opacity is taken from Ossenkopf {\em et al.}
(1992), with the standard MRN grain size distribution (Mathis, Rumpl \& Nordsiek,
1977). Dust density is described by $r^{-2}$ with an arbitrary inner envelope
radius, $r_1$. The outer envelope radius is
irrelevant as long as it is much larger than $r_1$, and we fix it at the value
of $3 \times 10^{17}$ cm. In the interrupted mass loss model, the inner envelope
radius increases with $t$, the time elapsed since the mass loss stopped as
\eq{
 r_1 = r_1(0) + {\rm v} \, t = r_1(0) + 3.2 \times 10^{13}  \,{\rm cm} \,\, v_{10} \, \tyr
}
where $t$ = \tyr\ years, the expansion velocity v = $v_{10}$ 10 km s$^{-1}$,
and $r_1(0)$ is the envelope inner radius at $t=0$. This radius depends
on the grain optical properties and dust condensation temperature,
and the stellar temperature and luminosity (IE97). For typical values
\eq{
     r_1(0) = 3 \times 10^{14} \,{\rm cm} \,\, L_4^{1/2},
}
where the stellar luminosity $L = L_4 10^4$ \Lo.

We specify $r_1$ by the temperature at the inner edge at time $t$, T$_1$.
This temperature decreases with time because the distance between the star
and the envelope inner edge increases. This relationship is approximately
given by (IE97, eq. 30)
\eq{
  {T_1 \over T_1(0)} = \left( {r_1(0) \over r_1} \right) ^{2 \over 4+\beta}
}
where $\beta$ is the power-law index describing the opacity wavelength
dependence. This expression is strictly correct only in the optically
thin regime, and furthermore, the silicate dust opacity cannot be
described by a simple power law. However, it provides a good description
of detailed model results with $\alpha = 2/(4+\beta) \about 0.40\pm0.05$.
Combining the above expressions gives
\eq{
{T_1 \over T_1(0)} =
    \left(1 + 0.1 {v_{10} \tyr \over L_4^{1/2}} \right) ^{-\alpha}.
}

As already indicated in \S2 (c.f. Figure 1), the distribution of SR stars
in the K-[12] vs. [12]-[25] color-color diagram implies a lower limit on
$T_1$ of \about 300 K. Assuming $T_1(0)$ = 800 K, $v_{10} = L_4 = 1$ implies
a time scale of 100 yr.

The second free parameter is the dust optical depth which we
specify as the visual optical depth, \tauV. This parameter controls
the amount of dust and increases with \Mdot, the mass-loss rate during the
high mass loss phase. The steady-state\footnote{The steady-state assumption 
does not strictly apply here. However, the crossing time for the acceleration
zone ($\about$ 2$r_1$, corresponding to $\about$10 yr) is much shorter than
the presumed duration of the high mass-loss rate phase, and thus eq.(5) can
be used to estimate approximate mass-loss rate.} radiatively driven wind 
models give for silicate dust (Elitzur \& Ivezi\'c 2000)
\eq{
 \Mdot = 2 \times 10^{-6} \, \Mo \, {\rm yr}^{-1}
 \,  \tauV(0)^{3/4} \, L_4^{3/4} \left( r_{gd} \over 200 \right) ^{1/2}
}
where $r_{gd}$ is the gas-to-dust ratio, and \tauV(0) is the optical
depth before the mass loss stops. Note that \Mdot\ $\propto \tauV^{3/4}$
rather than \Mdot\ $\propto$ \tauV\ due to dust drift effects.

After the mass loss stops, the optical depth decreases with time because
it is inversely proportional to $r_1$ due to the envelope dilution
\eq{
{\tauV \over \tauV(0)} = \left(1 + 0.1 {v_{10} \tyr \over L_4^{1/2}} \right) ^{-1}.
}

The models could be parametrized by \Mdot\ and $r_1$ (or $t$) instead of
\tauV\ and $T_1$. However, in such a case the parametrization of a model
spectrum would involve at least two additional parameters ($L$ and
$r_{gd}$) and would artifically introduce a four-dimensional fitting
problem. As the scaling properties of the radiative transfer imply (IE97),
the described model is only a two-dimensional problem fully
specified by \tauV\ and $T_1$. It is only the correspondence between
these two parameters and other quantities such as \Mdot\ that involves
additional assumptions about e.g. $r_{gd}$.

In the interrupted mass loss model the changes of $T_1$ and \tauV\ as the
envelope expands are correlated. Combining eqs. (4) and (6) gives
\eq{
      {T_1 \over T_1(0)} = \left( {\tauV \over \tauV(0)} \right) ^\alpha,
}
and thus during the expansion
\eq{
       T_1 \tauV^{-\alpha} = C = const.
}
where the value of constant C is determined by \Mdot\ during
the high mass loss phase. The model predicts that the distribution of
best-fit $T_1$ and \tauV\ should resemble a strip in the \tauV--$T_1$ plane whose
width is determined by the intrinsic distribution of the mass-loss rates during
the high mass-loss rate phase, and with the position along the strip parametrized
by the time since the mass loss stopped.

Another model prediction is for the $T_1$ distribution of sources in an
unbiased sample (or equivalently for the \tauV\ distribution since \tauV\
and $T_1$ are correlated). Eq. (4) shows that the first derivative of $T_1$
decreases with time and thus the number of observed sources should increase
as $T_1$ decreases. Assuming $dN(T_1) \propto dt(T_1)$, where $dN$ is
the number of sources in a given $T_1$ bin, the model implies
\eq{
{dN \over dT_1} \propto T_1 ^ {- {1+\alpha \over \alpha}} \propto T_1^{-3.5}.
}
This prediction applies for sources with $T_1$ smaller than the dust
condensation temperature, $T_c$. The number of sources with $T_1$ comparable
to $T_c$ depends on the duration of the high mass loss phase such that
the ratio of the number of sources with $T_1 < T_c$, to the number of sources
with $T_1 \about T_c$, is equal to the ratio of times spent in the low and
high mass-loss phases.

\begin{figure*}
\begin{minipage}{\textwidth}
\label{fig_seq}
\centering \leavevmode \psfig{file=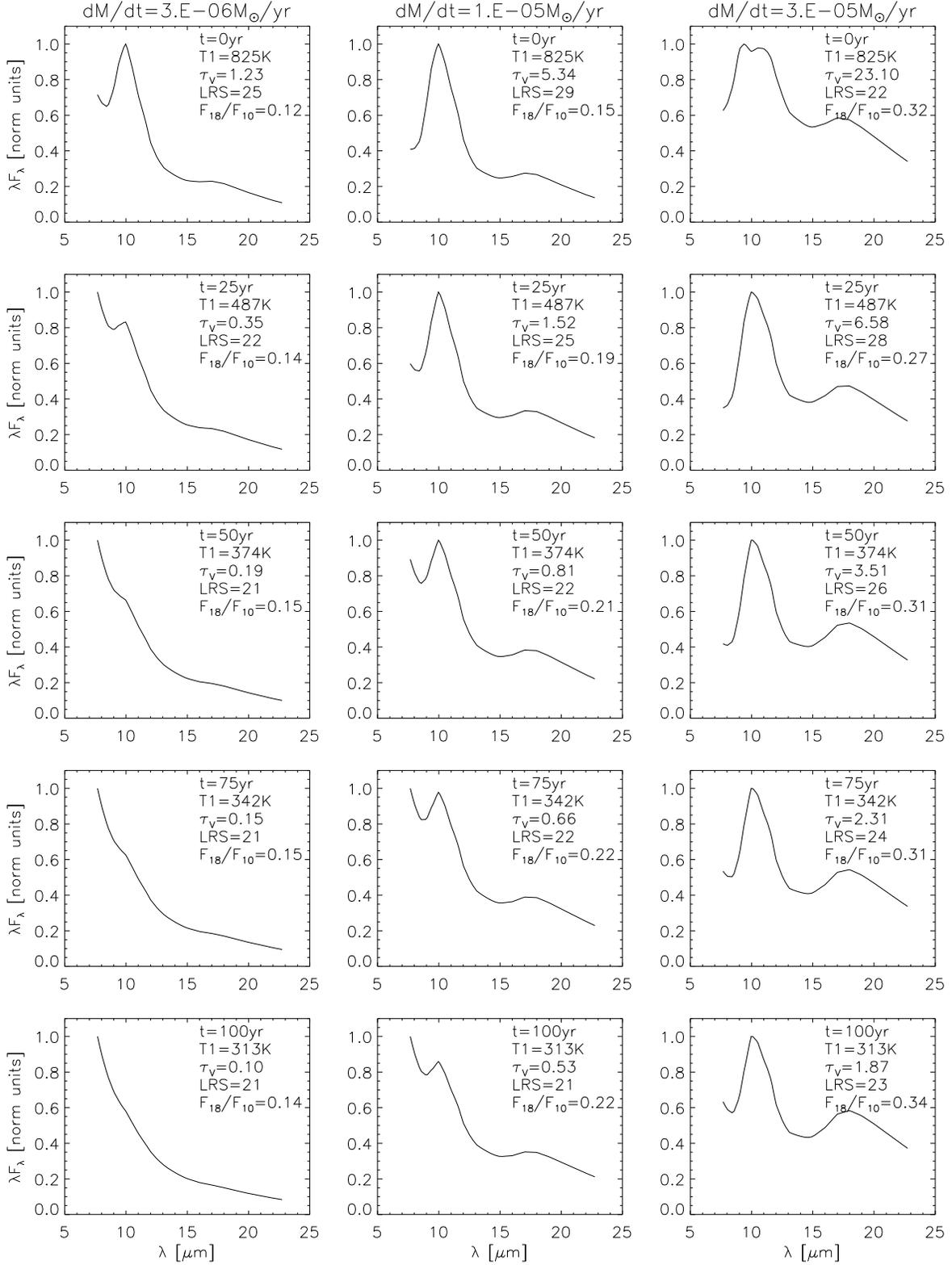,width=0.9\hsize,clip=}
\caption{An illustration of the two-dimensionality of LRS spectra.
Each column displays a time evolution of model spectra after the
mass loss interruption, calculated in steps of 25 years. The spectra in
3 columns are displayed for different mass-loss rates before the interruption,
increasing from left to right, and indicated on top of each column.
Each spectrum is fully characterized by
\tauV\ and $T_1$, which equivalently can be expressed as time $t$, marked
in each panel, and mass-loss rate.
The corresponding values of LRS class, which measures the 9.7 \mic\ silicate
feature strength, and the F$_{18}$/F$_{10}$ ratio are also marked in each
panel (see text for details).}
\end{minipage}
\end{figure*}

To summarize, the interrupted mass loss models assume a sudden
drop in the mass-loss rate after which the envelope freely expands.
The infrared spectrum emitted during this phase depends on only two
free parameters, the temperature at the inner envelope edge, $T_1$ and
the envelope optical depth \tauV. Both of these quantities decrease
with time in a correlated way described by eq. (8).

Note that the two free parameters imply a two-dimensional family of
model LRS spectra. It is usually assumed that the shape of an LRS spectrum
for silicate grains is fully specified by the 9.7 \mic\ silicate feature
strength. However, in this model even for a fixed 9.7 \mic\ silicate feature
strength, there is a family of spectra with differing shapes due to the
varying second parameter. We find that the ratio of fluxes at 18 \mic\
and 10 \mic, F$_{18}$/F$_{10}$, is a convenient observable, in addition to
the 9.7 \mic\ silicate feature strength, which can be used to parametrize
the LRS spectra. That is, each combination of $T_1$ and \tauV\ corresponds
to a unique combination of the 9.7 \mic\ silicate feature strength and
the F$_{18}$/F$_{10}$ ratio, and each such LRS spectrum corresponds to
a unique pair of $T_1$ and \tauV. Of course, assuming that all stars have
the same $L$, $r_{gd}$ and $v_{10}$, a given LRS spectrum corresponds to
a unique combination of \Mdot\ and $r_1$, and vice versa.

This two-dimensionality of LRS spectra is illustrated in Figure 6 which
displays a time evolution of model spectra after the mass loss interruption 
(for details see the figure caption). Each spectrum is fully characterized by
\tauV\ and $T_1$, which equivalently can be expressed as time $t$,  marked
in each panel, and mass-loss rate indicated on top of each column.
The corresponding values of LRS class, which measures the 9.7 \mic\ silicate
feature strength, and the F$_{18}$/F$_{10}$ ratio are also marked in each
panel. Note that some spectra have the same LRS class (e.g. $t$=0 yr in the
first column, $t$=25 yr in the second column, and $t$=50-75 yr in the
third column) and yet are clearly distinguishable both by their overall
shape and the values of the F$_{18}$/F$_{10}$ ratio. Another important
detail is that the spectra with comparatively large F$_{18}$/F$_{10}$ ratio
are obtained only for comparatively low $T_1$, as discussed earlier.

While we assume that the mass loss ceases completely, the data provide
only a lower limit on the mass-loss rate ratio in the two phases of
high and low mass loss.  We have explored a limited set of models where
the mass-loss rate drops for a factor 3, 5, 10, and 100, and found that
with the available data we cannot distinguish cases when the mass-loss
rate drops by more than a factor of 5. Thus, this is a lower limit
on the ratio of mass-loss rates in the two phases.

We have computed \about 2000 model spectra parametrized
on logarithmic grids with 62 \tauV\ steps in the range $10^{-3}$--350,
and 32 $T_1$ steps in the range 100 K -- 1400 K. Models calculated on
a finer grid in either \tauV\ or $T_1$ vary less between two adjacent
grid points than typical noise in the data. A subset of models with
$T_1$ restricted to the interval 600--1400 K is hereafter called ``hot-dust''
models. We separately fit these models to all sources in order to test
whether ``cold-dust'' models (with $T_1 <$ 600 K) are necessary to
improve the fits.

\subsection{              Fitting Technique           }

The best fit model for each source is found by using a
$\chi^2$ minimization routine applied to the source and model
spectra in the IRAS LRS spectral region (8 \mic -- 23 \mic).
The $\chi^2$ variable is defined as
\begin{equation}
\chi^2 = \frac{1}{N-2} \sum_{i=1}^N \frac{\left[
\lambda_i F^O_\lambda(\lambda_i) - \lambda_i F^M_\lambda(\lambda_i)
\right]^2}{\sigma_O^2 + \sigma_M^2},
\end{equation}
where $F^O_\lambda$ is the observed flux and $F^M_\lambda$ is the model
flux. The error $\sigma_O$ of the IRAS LRS data was
estimated as the root-mean-square (rms) difference between the raw and a
cubic spline smoothed LRS spectra. The model error $\sigma_M$ is
taken as the rms difference between the two closest models in the parameter
space. The $\chi^2$ variable was then normalized to the number of
wavelengths $N$, minus the number of fitting parameters ($\tau_V$ and $T_1$).
By comparing each source spectrum with all spectra in a given model set,
we determine the $\chi^2$ and select a model with the lowest $\chi^2$ as
the best-fit model. This procedure is repeated independently for the set
of all models, and for the ``hot dust'' subset.

\begin{figure}
\label{selftest}
\centering \leavevmode \psfig{file=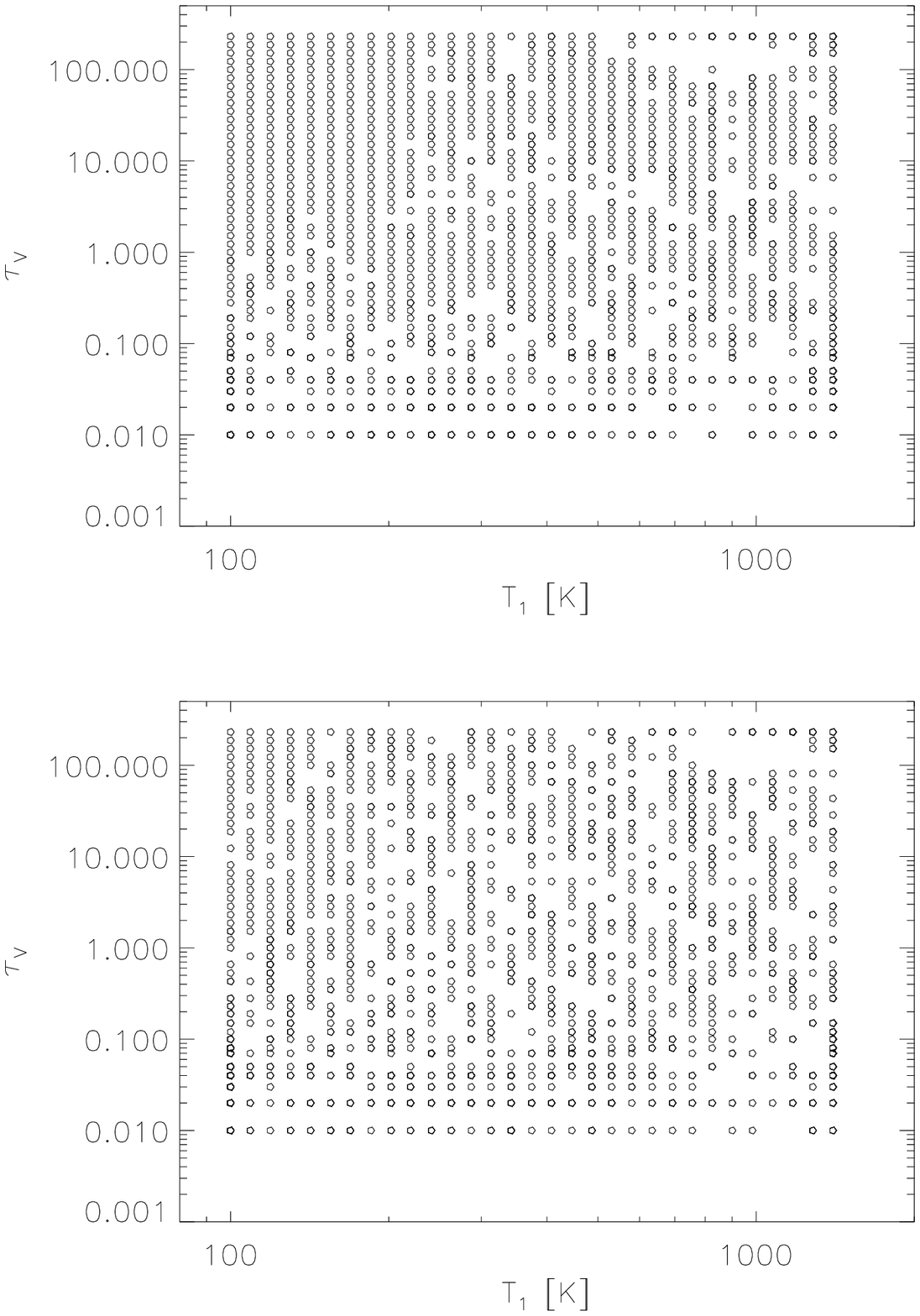,width=1.1\hsize,clip=}
\caption{Test diagram of the fitting procedure. Gaussian noise
with SNR = 10 (top) and 5 (bottom) is added to the full grid of models and
the resulting spectra are fitted with the noiseless spectra. Each circle
represent a successful recovery of the true \tauV\ and $T_1$, while
the missing points are incorrectly identified models. The majority of
``failed'' models fall on to a neighboring grid point.}
\end{figure}

As discussed in the previous section, the interrupted mass loss model
predicts a particular distribution of sources
in the \tauV--$T_1$ plane, which may be affected by hidden systematic biases
in the fitting procedure. In order to assess the level of such biasing, we
perform a ``self-fit" test for all models in the sample. We add Gaussian
noise similar to the noise in the LRS data to each of the DUSTY model spectra
and then pass such spectra through the fitting algorithm. We consider
two noise levels (signal-to-noise ratio, SNR), SNR = 10 typical for the
whole sample and SNR = 5 representative of the stars with the lowest-quality
LRS data.

The results are shown in Figure 7 where each circle represents a point
in the model grid; missing points are models for which the true
\tauV\ and $T_1$ were not recovered due to noise in the ``data". The majority
of such ``failed'' fits fall on to a neighboring grid point. The figure shows
that increasing noise can reduce the effectiveness of our procedure. However,
the $T_1$ distribution of the incorrectly identified models is uniform even in
the low SNR case, and thus within the limits of our sample the fitting procedure
does not introduce significant biasing in the best-fit parameters.

\subsection {            Best-fit Model Results          }

The quality of a fit is described by the best fit $\chi^2$
variable. A good fit, in which the details of the source spectra are
reproduced by the model (e.g. Figure 8, panel $a$), generally has $\chi^2
\lesssim 5$. Fits with $5 \lesssim \chi^2 \lesssim 10$ are still acceptable,
even though some secondary features in the source spectrum cannot be fully
reproduced by the model. This seems to be mainly
due to the inability of the adopted silicate opacity to describe all observed
spectral features (e.g. the 13 \mic\ feature).
A larger $\chi^2$ (\ga 10) indicates
more serious discrepancies in the fit, that can range from the convergence
errors (e.g. when the fitting routine cannot find a well defined minimum
in the $\chi^2$ surface) to poor fits indicating that the adopted
opacity and/or radial dust density law are grossly inadequate. Such sources
represent less than 10\% of the sample.

\subsubsection {          The Fit Quality         }

The statistics of the best-fit $\chi^2$ values for sources with
$\chi^2 < 10$ are shown in Table 3. The table reports the fraction of sources in
each subsample (Mira, NM, SRa, spSR, and lpSR) which cannot be fitted with
a $\chi^2$ less than 3 and 10 (``excellent" and ``acceptable" fits).
The sources are fitted with two sets of
models: ``all temperatures" models which include the full $T_1$ grid, and
the ``hot dust''
models with $T_1 > 600$~K. As evident, a larger fraction of sources in all
subsamples can be fitted within a specified $\chi^2$ limit with the ``all"
model set than with the ``hot dust" models. This is not surprising since
the former model set includes a greater variety of spectra due to the two free
parameters, as opposed to the restrictions on one free parameter in
the latter set.
Nevertheless, the improvement in the best fit quality is more significant
for the NM sample (and all NM subsamples) than for Mira stars. For example,
more than a half of NM stars cannot be fitted with a $\chi^2 < 3$ by using
``hot dust" models, while the inclusion of models with low $T_1$ decreases
their fraction to less than 20\%. At the same time, the corresponding fractions
for Mira stars are 42\% and 35\% indicating that low-$T_1$ models are not
necessary to model LRS data of these stars.

\begin{table}
\begin{center}

\begin{tabular}{lccccc}
\hline
Sample & $\chi^2$$\la$$3$   & $\chi^2$$\la 10$  &
       & $\chi^2$$\la$$3$   & $\chi^2$$\la 10$    \\
        & all & models  & & hot & dust  \\
\hline \hline
 Mira                        & 35\% &  3\%  & &  42\% & 12\% \\
 NM                          & 18\% &  1\%  & &  54\% & 22\% \\
 SRa                         & 21\% &  2\%  & &  33\% &  4\% \\
 spSR ($P < 100$\days)       & 18\% &  0\%  & &  67\% & 21\% \\
 lpSR ($P \gtrsim 100$\days) & 21\% &  2\%  & &  48\% &  6\% \\
\hline
\end{tabular}

\caption{The best fit statistics. The table lists the fraction of sources
in each subsample which cannot be fitted with the specified $\chi^2$ criteria.
The second and third column correspond to ``all'' (temperature) models, and 
the third and fourth column to ``hot dust" models (see text).}
\smallskip

\end{center}
\end{table}

An example of the improvement in the fit quality due to a lower $T_1$
is shown in Figure 8. The model spectra are plotted by dashed dotted lines
and the IRAS LRS data by solid lines. In panel $a$ the SRb star TY Dra is fitted
with a cold dust model ($T_1 \simeq 370$~K), resulting in a very small $\chi^2$.
When the dust temperature is limited to the range 600--1400~K, the fit is unable
to reproduce the source spectral energy distribution, as shown in panel $b$.
The panels $c$ and $d$ show good quality fits of a Mira star with relatively
hot dust ($T_1$ = 825~K) and an SR star with very low inner shell temperature
($T_1$ = 270~K).

\begin{figure*}
\begin{minipage}{\textwidth}
\label{examples}
\centering \leavevmode \psfig{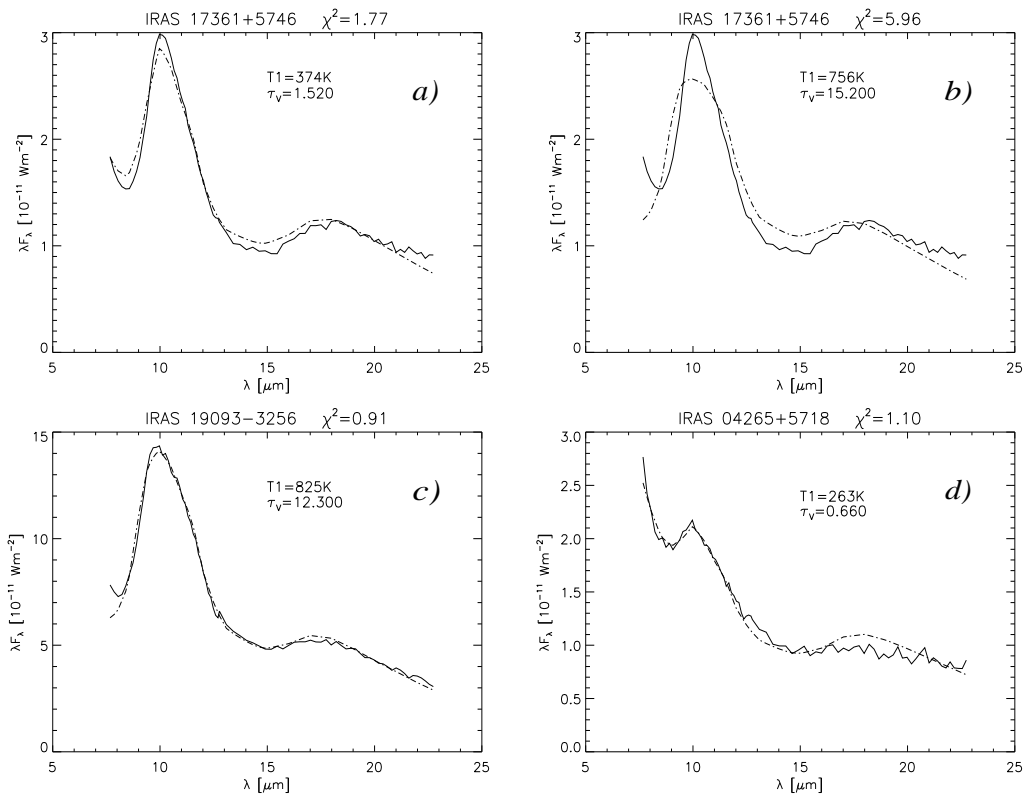}
\caption{An example of model fits (dashed dotted line) to the IRAS LRS data
(solid line) for Mira and NM sources. In panel $a$ the SRb star TY Dra is
fitted with a cold dust model ($T_1 \simeq 370$~K), resulting in a very
small $\chi^2$. When the dust temperature is limited to the range 640--1400~K,
the fit is unable to reproduce the source spectral energy distribution, as
shown in panel $b$. The panels $c$ and $d$ show good quality fits of a Mira
star with relatively hot dust ($T_1 = 825$~K) and an SR star with very
low inner shell temperature ($T_1 = 270$~K).}
\end{minipage}
\end{figure*}

\subsubsection { Distribution of Best-fit Parameters    }

The distribution of best-fit \tauV\ and $T_1$ is
shown in Figure 9. Sources from Mira and NM subsamples are marked
by solid and open circles, respectively (since the differences in
best-fit parameters between the various subsamples of NM stars
are not statistically significant, we consider only Mira/NM subsamples
hereafter). A small random offset (up to 1/3
of the parameter grid step) has been added to each $T_1$ and $\tau_V$, in
order to separate the sources with identical best fit parameters, which
would otherwise appear as a single point on the diagram. The sources are
distributed not randomly but rather along a diagonal strip. The density
of sources along the strip has a local minimum for $T_1$ \about 500 K
(\tauV \about 5) suggesting a division into four regions as shown by the
thin solid lines. The source counts in the quadrants, labeled
clockwise from I to IV, are given in Table 4. Although both Mira and NM
stars are present along the whole strip, Mira tend to aggregate in the
upper right quadrant with higher $T_1$ and \tauV. On the contrary, NM stars
aggregate in the lower left quadrant with $T_1 <$ 500 K and
$\tau_V <$ 0.5.

\begin{table}
\begin{center}

\begin{tabular}{lrrrr}
\hline
Sample               &   I &  II & III &  IV \\
\hline \hline
Mira                 &  59 &   1 &  32 &   1 \\
NM                   &  72 &   2 & 169 &   1 \\
SRa                  &  20 &   1 &  26 &   0 \\
\hline
\end{tabular}

\caption{The source distribution in $\tau_V$ vs. $T_1$ diagram (see Figure
9).}
\smallskip

\end{center}
\end{table}

\begin{figure*}
\begin{minipage}{\textwidth}
\label{tauVvsT1}
\centering \leavevmode \psfig{file=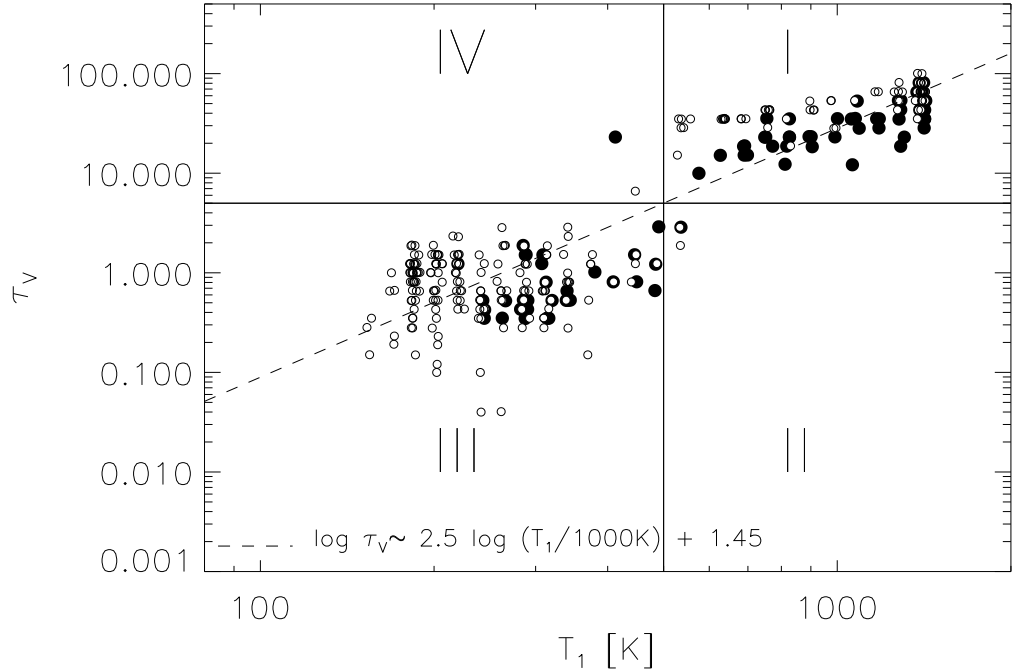,width=0.8\hsize,clip=}
\caption{The distribution of best-fit $\tau_V$ and $T_1$
for Mira stars (solid circles) and non-Mira stars (open circles).
A small random offset (up to 1/3 of the parameter grid step) has been
added to each $T_1$ and $\tau_V$, in order to separate the sources
with identical best fit parameters, which would otherwise appear as a single
point on the diagram. The plot is divided in four regions, according to the source
segregation; the counts in the four quadrants are given in Table 4.}
\end{minipage}
\end{figure*}

The strong correlation between the best-fit \tauV\ and $T_1$
seen in Figure 9 is probably not due to hidden biasing of the fitting algorithm
(c.f \S3.2). Indeed, such correlation is expected in the interrupted mass loss
model as discussed in \S3.1. Eq.(8) predicts
\tauV $\propto$ $T_1^{1/\alpha}$ = $T_1^{2.5}$
shown as a thick dashed line in Figure 9. The close agreement
between the source distribution and predicted correlation is evident.
Fitting a power law to the source distribution gives a best-fit power-law
index of 2.8 for Mira stars, and 2.5 for NM stars, consistent with the
model prediction, given the uncertainty of $\alpha$ ($\pm0.05$).

Another model prediction pertains to the distribution of best-fit $T_1$.
For sources in the quiescent phase, eq. (9) predicts $dN/dT_1$ $\propto$
$T_1^{-3.5}$, or equivalently, $dN/dlog(T_1)$ $\propto$ $T_1^{-2.5}$. The
histogram of best-fit $T_1$ is shown in Figure 10 for Mira stars (dashed line)
and NM stars (solid line). The relation $dN/d\log(T_1)$ $\propto$ $T_1^{-2.5}$
is shown as a dotted line,
normalized to the NM bin with $T_1$ $\approx$ 200 K.
It is evident that the $T_1$ distribution is very different for Mira and NM
stars. While this distribution is flat for Mira stars, with about 2/3 of stars
having $T_1 >$ 500 K, 2/3 of NM stars have $T_1 <$ 500 K. These stars closely
follow the model prediction for stars whose envelopes are freely expanding
after their mass loss has stopped.

The low-$T_1$ of the NM distribution in Figure 10 abruptly ends at
$T_1$ \about 200 K although the model grid extends to $T_1$ = 100 K.
As discussed in IK98, this sharp end may indicate that the mass loss
resumes after the quiescent phase because otherwise the envelopes with
100 K $< T_1 <$ 200 K should be detected. Eq. (4) implies that the time scale
for the envelope expansion until $T_1$ reaches such a low temperature
is of the order 100 years. If the mass loss starts again, in less than
10 years the new shell would occupy the region where the dust
temperature is higher than about 600 K, and thus about 10\% of stars
could be in this phase. The only difference
between such a double-shell envelope and a steady-state envelope with
smooth $r^{-2}$ dust density distribution is the lack of dust with
temperatures in the range 200--600 K. Detailed model spectra in the
spectral range 5-35 \mic\ for such
double-shell envelopes show that they are practically indistinguishable
from spectra obtained for envelopes with $r^{-2}$ dust density distribution.

The fraction of NM stars with $T_1 >$ 500 K is \about 1/3. This fraction
provides an upper limit for the duration of high mass loss phase of \about
50 years. However, it would not be surprising if this estimate is wrong
by perhaps a factor of 2 since the analysis presented here only provides
a lower limit on the ratio of mass-loss rates in the high and low mass loss
phases. Thus, the observations do not strongly exclude the possibility
that the duration of both phases may be the same.

\begin{figure}
\label{T1histo}
\centering \leavevmode \psfig{file=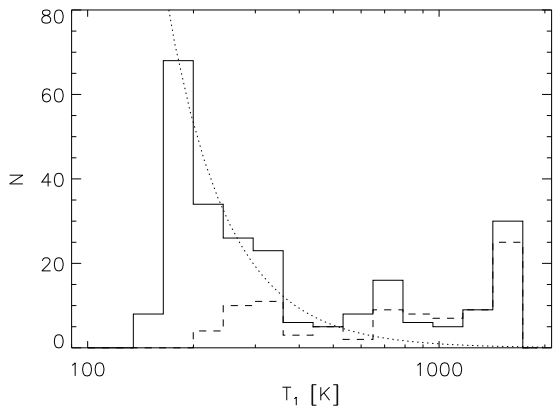,width=\hsize,clip=}
\caption{Histogram of best-fit $T_1$ for Mira stars (dashed line) and
NM stars (solid line). The dashed line shows relation dN/dlog($T_1$)
$\propto$  $T_1^{-2.5}$, normalized to the NM bin with $T_1 \approx$ 200 K.}
\end{figure}

\subsection {    The Distribution of Mass-loss Rates       }

The width of the strip formed by the source distribution in the
\tauV\ vs. $T_1$ diagram (Figure 9) is determined by the intrinsic
distribution of mass-loss rates. We calculate mass-loss rates from
the best-fit \tauV\ and $T_1$ by using
\eq{
  \Mdot = \tauV^{3/4} \left( 1000 \, {\rm K} \over T_1 \right)^{2.5}
          10^{-6}\, \Mo \,{\rm yr}^{-1}
}
which is derived by combining eqs. (5-7) and assuming
$(T_1(0)/1000 {\rm K})^{2.5} \, L_4^{3/4} \, (r_{gd}/200)^{1/2}= 0.5$.
Figure 11 shows the
histograms of calculated mass-loss rates for Mira stars (dashed line)
and NM stars (solid line). The histograms are rather narrow,
and furthermore, they are the same within the uncertainties, according
to F-test and t-test analysis. Such a similarity of mass-loss rates between
Mira and NM stars is somewhat surprising because a ``standard" result is
that NM stars have several times smaller mass-loss rates than Mira stars
(e.g. Habing 1996). A plausible explanation for the smaller mass-loss
rates usually derived for NM stars may be the lack of correction for
lower $T_1$ (c.f. eq. 11) when converting the envelope optical depth to
mass-loss rate. The results presented here indicate that the majority
of both Mira and NM stars have mass-loss rates of the order 10$^{-5}$.
An alternative explanation is that the KHc sample of NM stars is biased
towards high mass-loss rates, although such an effect is not expected
for their selection procedure. Additional selection criteria employed in
this work did not affect a sufficiently large number of NM sources to
result in such biasing. It is not possible to distinguish between those
two possibilites without a large and unbiased sample with uniformly measured
mass-loss rates by an IR-independent method (e.g. molecular emission).

\begin{figure}
\label{Mdothisto}
\centering \leavevmode \psfig{file=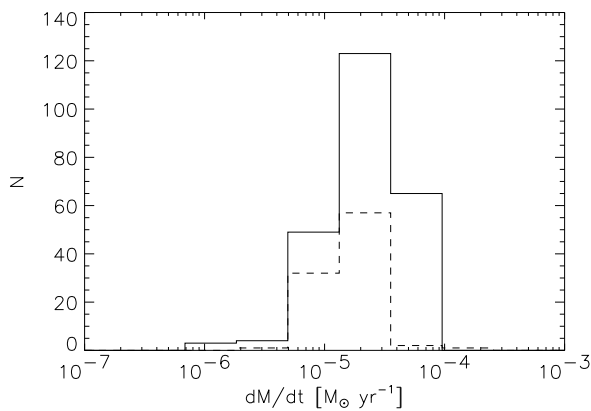,width=\hsize,clip=}
\caption{Histogram of derived mass-loss rates for Mira stars (dashed line)
and NM stars (solid line).}
\end{figure}

\section{ DISCUSSION }

The analysis presented here shows that the observed differences
in infrared emission from oxygen rich AGB stars of different
variability type are statistically
significant. Mira stars and NM stars clearly segregate
in the K-[12] vs. [12]-[25] color-color diagram, and also have
different detailed properties of their IRAS LRS spectra. We find no statistically
significant differences between various subsamples of NM stars. The
differences between Mira and NM stars are hard to interpret in the context
of steady state wind models. In particular, these models
do not provide satisfactory description of the infrared emission from the
majority of NM stars. Possible ways to augment these models
include changing the shape of stellar spectrum, employing grains with
altered absorption efficiency, or assuming that the majority of NM
stars have envelopes without hot dust ($\ga$ 500 K).

We find that the absence of hot dust for NMs is the most plausible way
to explain the observations, although the available
data cannot rule out the possibility that the differences in circumstellar 
dust chemistry produce the observed differences in infrared emission.
For example, Sloan {\em et al.} (1996) found that the ``13 micron'' feature 
occurs somewhat more frequently in SRb stars (75\%-90\%) than in all AGB stars 
with silicate dust (40\%-50\%). On the other hand, this difference could be
simply due to different densities in the dust formation region, as pointed
out by Speck {\em et al.} (2000). We are in the midst of a detailed
study exploring the model spectra obtained for a large sample of different
dust types and grain size distributions (Helva\c{c}i {\em et al.} 2001).

We provide detailed analysis of models where the absence of hot dust
for NM stars is interpreted as an interruption of the mass loss with time
scales of the order 100 years, and a drop in the mass-loss rate of at least a
factor of 5. In such a case, envelope freely expands and the dust temperature
decreases with time. This hypothesis predicts a correlation
\tauV $\propto T_1^{2.5}$ between the envelope optical depth, \tauV\
and the dust temperature at the inner envelope
edge, $T_1$, which we find consistent with the data. We find that the model
prediction for the distribution of best-fit $T_1$ also agrees with
the data, although this constraint may be plagued by hidden biases in the sample.
We interpret the sharp low-$T_1$ end of the source distribution as
evidence that the mass loss resumes and estimate the duration of high mass
loss phase to be of the same order as the duration of the low mass loss phase.

The number of Mira stars with low $T_1$ is about 1/3 and it is not clear
from the present analysis whether this represents evidence for the
interrupted mass loss. It was proposed by Kerschbaum et al. (1996), based
on similar number densities and scale heights, that AGB stars oscillate
between NM and Mira phases. This idea was further advanced by IK98
based on their analysis of the source distribution in the K-[12] vs.
[12]-[25] diagram. However, recent results by Lebtzer \& Hron (1999),
who compare the abundance of $^{99}$Tc in a large sample of SR stars and Mira
stars, seem to rule out this hypothesis. The $^{99}$Tc abundance is characterized
by a quick increase during the first thermal pulse, after which it presumably
stays constant (Busso {\em et al.} 1992). Consequently, a similar fraction
of Tc-rich stars should be found in the Mira and SR samples, contrary to the
observations by Lebtzer \& Hron who find that a fraction of stars with
detectable Tc is lower for SR stars (15\%) than for Mira stars (75 \%).

While the available observations are consistent with the interrupted
mass loss model, the possibility that observed differences between Mira
and NM stars are due to different dust optical properties can not
be excluded by considering photometric data alone.
The most obvious and direct test of the interrupted mass loss hypothesis
is mid-IR imaging. When T$_1$ decreases by a factor of two, the inner
envelope radius increases by about a factor of four. Such a difference
should be easily discernible for a few dozen candidate stars with largest
angular sizes, which are already within the reach of the Keck telescopes
(e.g. Monnier {\em et al.} 1998). Molecular line observations can also be
used to infer the gas radial density distribution because various lines
form at different radii. For example, SiO8-7 (347.3 GHz) line is expected
to form much closer to
the star than, say, CO3-2 (345.8 GHz) line. If SRb/Lb stars really lack
material in the inner envelope, they should have lower SiO8-7/CO3-2
intensity ratios than Miras, provided that the excitation mechanisms
are similar. Another, indirect, test can be made by employing mass-loss
rates determined in molecular, either thermal or maser observations.
For a given mass-loss rate, NM stars should have bluer 12-K color
because their dust optical depth becomes smaller as the envelope expands.
While this effect is not very pronounced (difference is about 0.2-0.3
magnitudes), a sufficiently large sample might provide statistically meaningful
results.

For a few stars there are available spatially resolved observations (c.f. \S1)
and they seem to indicate mass loss changes with short time scales ($<$ several
hundred years). If proven correct, the mass-loss variations on time scale of about
100 years would significantly change our understanding of the stellar
evolution on the AGB since the known time scales are either much
shorter (stellar pulsations, \about 1 year), or much longer (He flashes,
\about 10$^5$ years). However, the time dependent wind models (Winters 1998)
seem to produce mass-loss rate variations with time scales much longer that
the pulsation period which drives the mass loss (\about 1 year). These longer
time scales are not fully understood (see however Deguchi 1997) and seem
to result from the complex interplay between the pulsation and dust formation
and destruction mechanisms. While the theoretically obtained time scales
(5--10~years) are shorter than those implied by the observations, the recent
developments show that it may be possible to increase the model time scale
to \about 100 yr (Winters J.M., priv. comm.). It is not clear why would Mira
and non-Mira stars exhibit different behaviors, unless the coupling mechanisms
are very sensitive to the details of the pulsation mechanism (see e.g.
Mattei {\em et al.} 1997). 

Another possibility is that the 100-year mass-loss rate modulations are caused by
the luminosity variations on similar time scales (Sahai {\em et al.} 1998).
Recent work based on the analysis of stars observed by both the IRAS and
HIPPARCOS surveys (Knauer, Ivezi\'c \& Knapp 2000) shows that the extensive
mass loss on the AGB seems to require a threshold luminosity of \about 2000 \Lo.
If the luminosity of non-Mira stars oscillates around this value, then the
resulting mass-loss rate would be in agreement with the model assumptions
discussed here. Within this hypothesis Mira stars could have slightly larger
luminosities (for perhaps a factor of 2) and exhibit a steady mass loss, a
possibility that seems to be consistent with the P-L relation since Mira stars
have somewhat longer periods that SR stars (Whitelock 1986).

\section*{Acknowledgments}

We are grateful to Franz Kerschbaum for providing to us JHKLM data
without which this work would have not been possible. We thank
Martin Jan Winters, Mikako Matsuura, Mathias Steffen, Janet Mattei,
John Monnier, Joseph Hron, Moshe Elitzur, Maia Nenkova, Dejan 
Vinkovi\' c and Mustafa Helva\c ci for illuminating discussions. 
We also thank the referee, Angela Speck, for comments which helped us 
improve the presentation. This work was partially supported by NSF grant 
AST96-18503 to Princeton University.

\label{lastpage}
\end{document}